# Element retrieval using Namespace Based on keyword search over XML Documents

Yang Wang
School of Software, Dalian University of Technology, Dalian, Liaoning 116620, P.R.C
wayag2000@yahoo.com.cn

Zhikui Chen
School of Software, Dalian University of Technology, Dalian, Liaoning 116620, P.R.C
zkchen@dlut.edu.cn

Xiaodi Huang
School of Computing and Mathematics of Charles Sturt University, Australia
xhuang@csu.edu.au

## ABSTRACT

Querying over XML elements using keyword search is steadily gaining popularity. The traditional similarity measure is widely employed in order to effectively retrieve various XML documents. A number of authors have already proposed different similarity-measure methods that take advantage of the structure and content of XML documents. They do not, however, consider the similarity between latent semantic information of element texts and that of keywords in a query. Although many algorithms on XML element search are available, some of them have the high computational complexity due to searching a huge number of elements. In this paper, we propose a new algorithm that makes use of the semantic similarity between elements instead of between entire XML documents, considering not only the structure and content of an XML document, but also semantic information of namespaces in elements. We compare our algorithm with the three other algorithms by testing on the real datasets. The experiments have demonstrated that our proposed method is able to improve the query accuracy, as well as to reduce the running time.

## Categories and Subject Descriptors

H.3 [Information Storage and Retrieval]: H.3.3 Information Search and Retrieval -- Search process; H.3.4 Systems and Software -- Performance evaluation (efficiency and effectiveness)

## General Terms

Algorithms

## 1. INTRODUCTION

Keyword search querying over XML elements has emerged as one of the most effective paradigms of information retrieval. To identify relevant results for an XML keyword query, different approaches lead to various search results in general. Some authors calculated the similarity between the content of XML documents and query only analyzing content and structure of XML(e.g., [13,29,20]). Multiple of Algorithms calculate the degree of text of elements matching with the keyword to produce the ranked result-list(e.g., DIL Query processing algorithm [4] and Top-k algorithm [22]). The classical method focus on TF-IEF formula to calculate the cosine similarity between elements and query(e.g., Tae-Soon Kim et al. [6]; Maria Izabel M et al. [17]; Zhang Yun-tao et al. [18];).

In particular, overlaps of elements in XML documents must be considered. From several overlapping relevant elements, we have to choose which one should be avoided to ensure that users do not get to see the same information for several times. Su Cheng Haw et al. [11] presented the TwigINLAB algorithm to improve XML Query processing. In this paper, we modify it to deal with the elements overlap occurring in keyword search results.

On the basis of previous work, we make the following contributions in this paper. We utilize the semantic information of namespaces in elements to filter the relevant components since the text of elements are commonly related with semantic information of namespace. The precision and recall of our algorithm shows that the non-text matching but semantic relevant elements with respect to the keyword can be effectively retrieved. Compared with traditional work, our algorithm also shows the better performance on time execution over a large collection of elements.

The rest of this paper is organized as follows: Section 2 introduces the element-rank schema by keyword search. Section 3 presents the Namespace Filter Algorithm (NFA). The experiments on the comparison of NFA and related methods are reported in Section 4. Related work is presented in Section 5, followed by the conclusion.

## 2. ELEMENT-RANK SCHEMA

In this section, we utilize the namespace of elements to describe our element rank schema. Another goal of utilizing namespace is to filter the relevant elements with the keyword in a query to reduce time execution compared with traditional algorithms.

Interestingly, namespaces can distinguish different elements containing the same markup that refers to different semantic meanings. As an illustration, consider two elements with the same markup of <table>:

*<table>*

  *<td>apple</td>*

  *<td>banana<td>*

*</table>*

*<table>*

  *<name>coffee table</name>*

  *<width>80</width>*

*</table>*

This will lead to the confliction when they are in the same XML document. Thus, we utilize different namespaces of 'h' and 'f' to distinguish them as below.

*<h:table xmlns:h = "http://.../fruit">*

      *<h:td>apple</h:td>*

      *<h:td>banana</h:td>*

*</h:table>*

*<f:table xmlns:f = "http://.../furniture">*

      *<f:name>coffee table</f:name>*

      *<f:width>80</f:width>*

      *<f:length>120</f:length>*

*</f:table>*

As discussed before, the text of elements is commonly related to the semantic information of their namespaces. Given the semantic information of namespace that is irrelevant to the keyword, it is not desirable to access all the elements containing this namespace. In order to calculate the semantic similarity between namespaces and keywords, we map semantic information of namespaces and keywords into different vectors in a concept vector space created by Singular Value Decomposition (SVD) in [8] over a collection of elements. In order to do this, Definitions 1 and 2 are provided as follows.

**Definition 1**: $prefix(v)$ :a function that maps the namespace of element $v$ into a vector and represents a special meaning in the concept vector space created by SVD.

**Definition 2**: $correlation(prefix(v), keyword)$ :the degree of relevance calculated by the cosine similarity between the namespaces vector of element v and the keyword vector in concept vector space created by SVD.

The value of the correlation is commonly normalized between the range [-1,1]. If the semantic meaning of namespaces is very close to that of the keywords, the value of '*correlation*' will be around 1. Nobert Govert et al. [29] proposed the concept of degree of relevance between elements. We extend it to include several intervals in [-1,1] to describe the degree of the semantic similarity of names- paces and keywords. Without loss of generality, Some definitions are provided to describe the degree of relevance between names- paces and keywords:

**Definition 3: High relevance:** the high correlation between namespaces and keywords which satisfies

$$\lambda_1 \leq correlation(prefix(v), keyword) \leq 1 \qquad (1)$$

**Definition 4: Common relevance:** the median correlation between namespaces and keywords which satisfies

$$\lambda_2 \leq correlation(prefix(v), keyword) < \lambda_1 \qquad (2)$$

**Definition 5: Irrelevance:** the lower correlation between namespaces and keywords which satisfies

$$-1 \leq correlation(prefix(v), keyword) < \lambda_2 \qquad (3)$$

In the above equations, we have $0 \leq \lambda_2 \leq \lambda_1 \leq 1$ ., Our ranking algorithm accesses the elements containing the namespaces that satisfy Eq.(1) or Eq.(2) rather than Eq.(3).

## 3.    Namespacec Filter Algorithm

In this section, we introduce some preliminary knowledge, followed by presenting our algorithm called the Namespace Filter Algorithm (NFA).

### 3.1    Preliminaries

The $tf-idf$ weight is commonly used to calculate the term weight in documents in the field of traditional information retrieval. The purpose of this work is to retrieve the appropriate nested elements that contain the relevant text to keywords instead of entire XML documents. So we extend $tf-idf$ to $tf_{t,e}-ief$ in order to tailor to elements in XML documents.

### Notations:

$tf_{t,e}$ the number of times that keyword t occurs in the text of element e.

$tf_{t,q}$ the number of times that keyword t occurs in the query q.

$ief = \log_{10} \dfrac{N}{ef}$ where $N$ is the total number of elements over a

collection of XML documents, and $ef$ is the number of elements that contain the keyword. We then give Definition 6 as below.

**Definition 6: keyword weights in elements and query**

$$W_{t,e} = \begin{cases} (1 + \log_{10} tf_{t,e}) \times ief & if \quad tf_{t,e} > 0 \\ 0 & otherwise \end{cases} \qquad (4)$$

$$W_{t,q} = (1 + \log_{10} tf_{t,q}) \times ief \qquad (5)$$

where $w_{t,e}$ is keyword weight in the text of element $e$ , and $w_{t,q}$ keyword weight in a query $q$ . We calculate the cosine similarity between query vector $\vec{q}$ and element vector $\vec{e}$ in Eq.(6) on text matching factor.

$$score(q,e) = \frac{\vec{q} \bullet \vec{e}}{\|q\| \|e\|} = \frac{\Sigma_{i=1}^{n} q_i e_i}{\sqrt{\Sigma_{i=1}^{n} q_i^2} \sqrt{\Sigma_{i=1}^{n} e_i^2}} \qquad (6)$$

where $e_i$ is the $i$th keyword weight in $\vec{e}$, and $q_i$ is the $i$th keyword weight in $\vec{q}$. Their weight values are calculated using Eq.(4) and Eq.(5), respectively.

In XML documents, elements are of varying size and nested. As relevant elements can be at any level of granularity, either an element or its children can be relevant to a given query. These facts commonly lead to a problem that the same resulting elements of a query based on keyword search will be presented to users for several times. As an illustration, Consider the structure of an XML document is shown as the labeled tree in Fig.1.

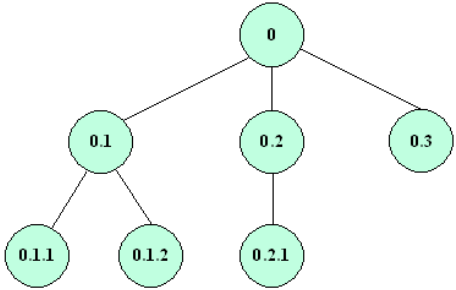

**Figure 1: Example of Elements with the label ID in the XML document tree**

Suppose the relevant element list after keyword search are listed in Table 1.

**Table 1:Example of ranked list**

| Rank | Self | Parent |
|------|------|--------|
| 1 | 0.2.1 | 0.2 |
| 2 | 0.2 | Root |
| 3 | 0.1 | Root |
| 4 | 0.1.1 | 0.1 |
| 5 | 0.1.2 | 0.1 |

Elements with ID 0.2.1 and 0.2 are overlapping, so are with 0.1, 0.1.1, 0.12. If one element's parent is the component of another element, the two relevant components can be merged into one. An element will be merged into its parent only if the number of the keyword occurring in this particular element is less than that of its parent element. In this way, there will be no overlap in the resulting list shown in Table 2.

**Table 2: Result list without overlap**

| Rank | Self | Parent |
|------|------|--------|
| 1 | 0.2.1 | 0.2 |
| 2 | 0.1 | Root |

Furthermore, we denote value[v] calculated by NFA in Section 3.2 as element v. Combining with Definition 2. The final comprehensive evaluation formula about relevant elements ranking is given as Eq.(7).

$$rank(v) = a_1 \times correlation(prefix(v), keyword) + a_2 \times value[v] \qquad (7)$$

where $a_1 + a_2 = 1$. In order to highlight the factor of namespace's semantic, we have $0 \le a_2 \le a_1 \le 1$

## 3.2 NFA description

In the following discussion, we will focus on presenting the Namespace Filter Algorithm(NFA) and how it performs based on the keyword search over a collection of elements.

Let A be a set consisting of different elements to be accessed by NFA, and the namespaces of elements in set A satisfy Eq.(1) or Eq.(2). Other elements not included in A will be neglected by NFA. The length[e] in Eq.(6) is defined as $\sqrt{\sum_{i=1}^{n} e_i^2}$

| **NFA : retrieve the ranked element based on the keyword** |
|---|
| **Input:query, a collection of relevant elements denoted as A** |
| **Output: top k elements of ranked result list** |
| Description: |
| 01    float value[N] = 0//N is the number of elements $\in$ A |
| 02    float Length[N] |
| 03    for each keyword t in the query |
| 04      do for each pair(element $\in$ A,tf(t,e)) |
| 05        do value[e] += $W_{t,e} \times W_{t,q}$ //Eqs.(4) and (5) |
| 06      end-for |
| 07    end-for |
| 08    for each element e |
| 09      do value[e] = value[e] / length[e] |
| 10    end-for |
| 11    merge the overlap |
| 12    calculate the rank[] with Eq.(7) |
| 13    return top K elements of rank[] over all documents |

**Figure 2:Namespace Filter Algorithm**

Value[e] in Fig.2 gives the degree of text matching between the text of element and keywords.

## 3.3 An Example

To evaluate the effectiveness of NFA, using an example, we perform it with different pair values of $\lambda_1$ and $\lambda_2$ in Eqs.(1) and (2) . We empirically provide an XML document named as record.xml in Fig.3 that consists of many elements with namespace 'c' with semantic "computer" and 'n' with "joy". Let the query be "data and space in algorithm", and set $\lambda_1$ in Eq.(1) and $\lambda_2$ in Eq.(2) to 0.8 and 0.6 ,respectively. SVD is commonly applied to documents in traditional information retrieval. We extend it to elements in this example.

<root1>

    <c:cs xmlns:c = "http://....../computer">

        <c:DBMS>

```
        <c:DB>attribute</c:DB>
        <c:DB>Management</c:DB>
      </c:DBMS>
  <c:programming>
      <c:complexity>data and space</c:complexity>
      <c:time>data in computer's Algorithm</c:time>
  </c:programming>
  <c:java>data of Algorithm in computer science</c:java>
  </c:cs>
  <n:joy xmlns:n = "http://....../happiness">
      <n:entertainment>
        <n:in>no space with audience's joy</n:in>
        <n:out>jackson  dance in large space</n:out>
      </n:entertainment>
  </n:joy>
</root1>
```

**Figure 3:Example of record.xml**

Each element in record.xml corresponds to a node in the tree with labeled ID in Fig.4.

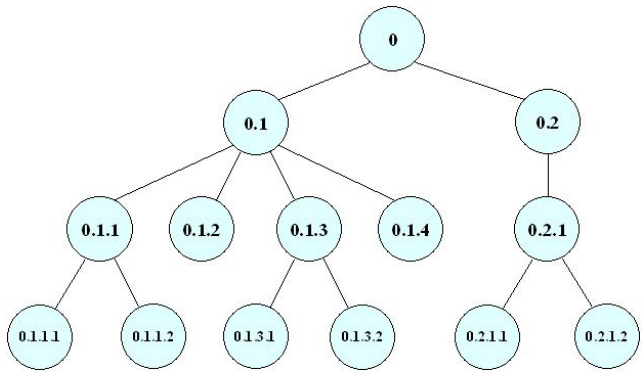

**Figure 4:Tree structure of record.xml**

Given the correlation value between the semantic meaning of namespaces: 'c', and 'n' , and that of the keywords :"data","space", and "Algorithm", we construct a term-element matrix denoted as M, the elements of which are term frequencies occurring over all of elements in record.xml in Fig.5.

|          | 0.1.1.1 | 0.1.1.2 | 0.1.2 | 0.1.3.1 | 0.1.3.2 | 0.1.4 | 0.2.1.1 | 0.2.1.2 |
|----------|---------|---------|-------|---------|---------|-------|---------|---------|
| Computer | 0       | 0       | 1     | 0       | 1       | 1     | 0       | 0       |
| data     | 0       | 0       | 1     | 1       | 1       | 1     | 0       | 0       |
| space    | 0       | 0       | 1     | 1       | 1       | 0     | 1       | 1       |
| Algorithm| 0       | 0       | 0     | 0       | 1       | 1     | 0       | 0       |
| joy      | 0       | 0       | 0     | 0       | 0       | 0     | 1       | 0       |

**Figure 5: Term-Element matrix M**

We normalize matrix M denoted as M1 in Fig.6.

$$\begin{bmatrix} 0 & 0 & 0.5774 & 0 & 0.5774 & 0.5774 & 0 & 0 \\ 0 & 0 & 0.5774 & 0.7071 & 0.5774 & 0.5774 & 0 & 0 \\ 0 & 0 & 0.5774 & 0.7071 & 0 & 0 & 0.7071 & 1 \\ 0 & 0 & 0 & 0 & 0.5774 & 0.5774 & 0 & 0 \\ 0 & 0 & 0 & 0 & 0 & 0 & 0.7071 & 0 \end{bmatrix}$$

**Figure 6:Matrix M1 that normalizes M**

M1 is decomposed into following three matrixes by SVD in Fig.7.

$$U = \begin{bmatrix} -0.4050 & 0.4285 & -0.1839 & 0.4256 & 0.6614 \\ -0.5874 & 0.3361 & 0.3163 & -0.6617 & -0.0637 \\ -0.6474 & -0.6863 & 0.1155 & 0.2915 & -0.1073 \\ -0.2435 & 0.4147 & -0.3754 & 0.3172 & -0.7261 \\ -0.1122 & -0.2458 & -0.8437 & -0.4420 & 0.1403 \end{bmatrix}$$

$$S = \begin{bmatrix} 1.8397 & 0 & 0 & 0 & 0 & 0 & 0 & 0 \\ 0 & 1.3770 & 0 & 0 & 0 & 0 & 0 & 0 \\ 0 & 0 & 0.6569 & 0 & 0 & 0 & 0 & 0 \\ 0 & 0 & 0 & 0.4126 & 0 & 0 & 0 & 0 \\ 0 & 0 & 0 & 0 & 0.3433 & 0 & 0 & 0 \end{bmatrix}$$

$$V = \begin{bmatrix} 0 & 0 & 0 & 0 & 0.6428 & 0.2843 & -0.7113 \\ 0 & 0 & 0 & 0 & -0.2946 & -0.7654 & -0.5722 \\ 0.5146 & -0.0328 & -0.2179 & -0.0774 & -0.8250 & & \\ 0.4746 & 0.1798 & -0.4647 & 0.6345 & 0.3520 & 0 & \\ 0.3878 & -0.4945 & 0.2135 & -0.1133 & 0.2159 & 0.5 & -0.4082 & 0.2887 \\ 0.3878 & -0.4945 & 0.2135 & -0.1133 & 0.2159 & -0.5 & 0.4082 & -0.2877 \\ 0.2920 & 0.4786 & 0.7839 & 0.2579 & -0.0681 & 0 & & \\ 0.3519 & 0.4984 & -0.1757 & -0.7066 & 0.3124 & 0 & & \end{bmatrix}$$

**Figure 7: The result of SVD(M1)**

In the following, we consider the reduced semantic space with two most informative dimensions. Let U1 be first two columns of U, S1 be the diagonal square matrix that contains the first two biggest eigenvalues 1.8397,13770 of S as diagonal elements, and other elements in S1are 0.V1 be the transpose of first two columns of V. We then build up a new term-element matrix M2 by using U1*S1*V1 in Fig.8.

$$\begin{bmatrix} 0 & 0 & 0.4024 & 0.2472 & 0.5804 & 0.5804 & -0.0650 & -0.0321 \\ 0 & 0 & 0.5707 & 0.4292 & 0.6475 & 0.6475 & 0.0937 & 0.1492 \\ 0 & 0 & 0.5813 & 0.7346 & -0.0059 & -0.0059 & 0.7997 & 0.8897 \\ 0 & 0 & 0.2490 & 0.1097 & 0.4559 & 0.4559 & -0.1426 & -0.1271 \\ 0 & 0 & 0.0950 & 0.1587 & -0.0874 & -0.0874 & 0.2222 & 0.2413 \end{bmatrix}$$

**Figure 8:New Term-Element matrix M2**

The correlation values between terms are shown in Fig.9a. We then normalize these values in Fig.9a to the range of [-1,1] as given in Fig.9b.

**a**

$$\begin{bmatrix} 2.7859 & 3.1450 & 0.9979 & 2.0658 & -0.0670 \\ 3.1450 & 3.7737 & 2.0238 & 2.1446 & 0.2023 \\ 0.9979 & 2.0238 & 3.9630 & -0.0134 & 1.0930 \\ 2.0658 & 2.1446 & -0.0134 & 1.6892 & -0.2831 \\ -0.0670 & 0.2023 & 1.0930 & -0.2831 & 0.3476 \end{bmatrix}$$

**b**

$$\begin{bmatrix} 0.8844 & 0.9984 & 0.3168 & 0.6558 & -0.0213 \\ 0.9984 & 1.1980 & 0.6425 & 0.6808 & 0.0642 \\ 0.3168 & 0.6425 & 1.2581 & -0.0043 & 0.3470 \\ 0.6558 & 0.6808 & -0.0043 & 0.5363 & -0.0899 \\ -0.0213 & 0.0642 & 0.3470 & -0.0899 & 0.1103 \end{bmatrix}$$

**Figure 9: Correlation value between different pair of terms in record.xml**

According to Fig.9b, the correlation values between the semantic meanings of namespaces 'c','n' and those of the keywords in the query are given in Table 3.

**Table 3:Correlation value between semantic of namespace 'c' ,'n' vectors and other three keyword vectors over elements in record.xml**

| Correlation | data | space | Algorithm |
|---|---|---|---|
| **computer** | 0.9984 | 0.3168 | 0.6558 |
| **joy** | 0.0642 | 0.3470 | -0.0899 |

Consider the keyword search for "data" or "Algorithm" in a query. As shown in Table 3, both the valued of correlation of Namespace 'c' vector with "data" and "Algorithm" vector satisfies Eq.(1). In contrast, the correlation value of Namespace 'n' vector and keyword vectors does not satisfy Eq.(1) and Eq.(2). So we have $\{0.2, 0.2.1, 0.2.1.1, 0.2.1.2\} \not\subset A$. The parameter values (in Section 3. 1) of elements in set A are listed in Table 4.

**Table 4:Times of "data","space","algorithm" occurring in query and revelant elements of record.xml**

| Dewey ID | $tf_{t,e}(data)$ | $tf_{t,e}(space)$ | $tf_{t,e}(algorithm)$ |
|---|---|---|---|
| 0.1 | 3 | 3 | 2 |
| 0.1.2 | 1 | 1 | 0 |
| 0.1.3 | 1 | 1 | 1 |
| 0.1.3.1 | 1 | 1 | 0 |
| 0.1.3.2 | 0 | 0 | 1 |
| 0.1.4 | 1 | 1 | 1 |

From line 05 to 11 of NFA in Fig.2, combining with Table 4, the value[e] s of elements in set A are shown in Table 5.

**Table 5: The ranked result-list with element overlap**

| Rank | Self | Parent | Value[e] |
|---|---|---|---|
| 1 | 0.1 | Root | 1.6160 |
| 2 | 0.1.4 | 0.1 | 1.5779 |
| 3 | 0.1.3 | 0.1 | 1.5779 |
| 4 | 0.1.2 | 0.1 | 1.4145 |
| 5 | 0.1.3.1 | 0.1.3 | 1.4145 |
| 6 | 0.1.3.2 | 0.1.3 | 1 |

As shown in Table 5, there exists the overlap between element 0.1 and other elements. After merging the overlap, the result is 0.1 including its descendent elements 0.1.2,0.1.3,0.1.4 as a whole components. The other resulting element is 0.1.1 including all its descendent elements. Let $a_1$ , and $a_2$ in Eq.(7) be 0.9 and 0.1,respectively, and the correlation value between namespace and keywords be 0.8271, which is the average correlation value between "data" and "Algorithm". Then we can get the final ranked result in Table 6.

**Table 6:Comprehensive ranking by Eq.(7)**

| Rank | Dewey ID | Score |
|---|---|---|
| 1 | 0.1 | 0.9060 |
| 2 | 0.1.1 | 0.7444 |

The search result is shown in Fig10.

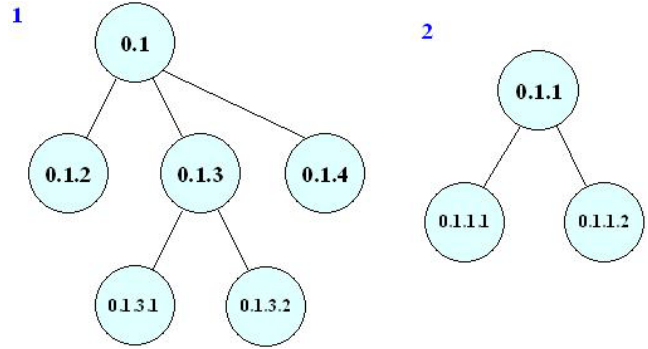

**Figure 10:Experimental result elements in record.xml retrieved by NFA with $\lambda_1$ in Eq.(1) be 0.8 and $\lambda_2$ in Eq.(2) be 0.6**

In order to exploit the relation between $\lambda_1$ in Eq.(1), $\lambda_2$ in Eq.(2) and search result, we assign different pair values to $\lambda_1$ and $\lambda_2$ such as 0.6 and 0.3. We still give the same query to perform NFA over record.xml. This time we focus on "space" in the query rather than "Algorithm" and "data". Table 3 shows that the correlation is 0.3, which satisfies Eq.(2) and namespace 'n' vector and "space" vector satisfy Eq.(1). Consequently, the search result performed by NFA is given in Fig.11.

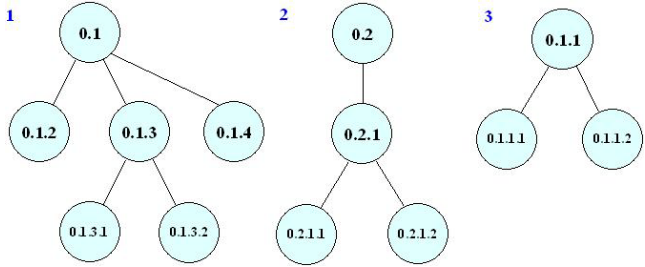

**Figure 11: Experimental result elements in record.xml retrieved by the Query with the condition of $\lambda_1$ in Eq.(1) and $\lambda_2$ in Eq.(2) be 0.6 and 0.3**

As shown in Figs. 10 and 11, the different degrees of semantic information relevance between the namespaces and keywords will lead to various search results by using NFA.

In summary, the degree of semantic relevance between the namespace and keywords depends not only on their semantic information similarity, but also on user-specified weights on other factors.

# 4. Experiments

In our experiments, we compare NFA with other related algorithms and methods on two metrics: precision and recall. The result of comparing NFA with the methods that have the similar Precision and Recall on aspect of time execution of algorithm is also presented. We set 0.9 to $\lambda_1$, 0.6 to $\lambda_2$, and 0.9 to $a_1$, 0.1 to $a_2$ in Eq.(7) to perform NFA.

## 4.1 Experimental Setup and results

**Equipment:** Our experiments are performed on a PC with a 2.33GHz Intel(R) Core(TM) 2 Duo CPU, 3.25 GB memory, and Microsoft Windows XP. The TermJoin algorithm [7], semantic tree creation algorithm [30], and NFA are all implemented in C++.

**Data set:** We have tested NFA on two data sets called Dataset1 [33] and Dataset2 [32], respectively. In order to show its performance, we add some namespaces to elements. Each namespace represents the general idea of text embedded in elements [33].

**Query set:** the query set consists of two parts with 13 queries that represent all kinds of queries over Dataset1 and Dataset2 in Table 7.

### Table 7:Query set on Dataset1 and Dataset2

**Dataset1**

Q11: pitch step B and octave 2

Q12: natural type

Q13: voice 1 and type eighth

Q14: music with voice 1 staff 1

Q15: music with beam begin and down

Q16: $16^{th}$ type in music

Q17: $16^{th}$ type and type of beam

Q18: $16^{th}$ type and duration 2

**Dataset2**

Q21: best table in furniture

Q22: best fruit table in furniture

Q23: eat apple at the table

Q24: have coffee at the table

Q25: the list of table

### 4.1.1 Precision and Recall

Precision is defined as the number of relevant elements retrieved by keyword search divided by the total number of elements, while recall refers to as the number of relevant elements retrieved by keyword search divided by the total number of existing relevant elements. We compare the precision and recall of NFA with the Termjoin algorithm [7], semantic tree creation algorithm [30] on Dataset1 and CAS Query [17] on Dataset2. We then calculate the precision and recall of top 20 components retrieved by each algorithm as reported in Fig.12.

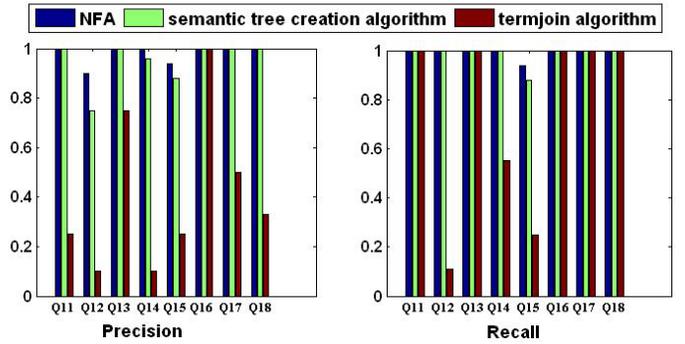

**Figure 12:Precision and Recall on Dataset1**

As shown in Fig.12, the term-join algorithm retrieves the relevant elements. However, it also retrieves some non-relevant elements. The basic idea of the Term-join algorithm is to calculate the degree of text matching of elements with keyword rather than the latent semantic information of text of elements. Both NFA and semantic tree creation algorithm efficiently solve the semantic information similarity between text of elements and keyword . However, they do not have the equal running time as given in Section 4.1.2. In [6,17], authors provide the methods that utilize the semantic information of markups in elements to calculate the semantic information similarity between the elements and query. However, sometimes it can only get the relevant components with various markups. In order to present the difference of search result of CAS Query [17] and NFA, we test both of them on Dataset2 consisting of elements with namespace 'h' and 'f' nested in the same markup <table>. Both of them are tested by queries from Q21 to Q25 in Table 7 and the precision and recall is shown in Fig.13.

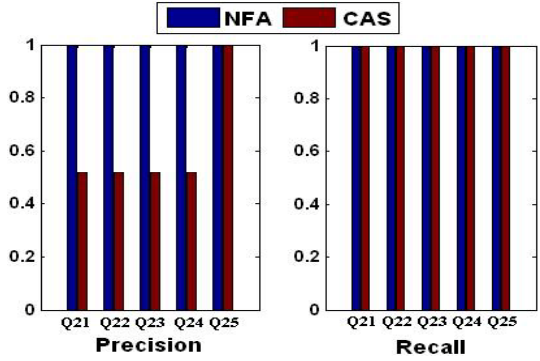

**Figure 13:Precision and Recall on Dataset2**

As discussed in Section 2.1, namespace can distinguish different elements even with same markup which leads to different precision of NFA and CAS in Fig.13.

### 4.1.2 Running time of NFA and Semantic tree creation Algorithm

In terms of running time in practice, we compare NFA and semantic tree creation algorithm. We test both of them on Dataset1, and plot the average running time based on the queries from Q11 to Q18 in Table 7 over 100 thousand elements from Dataset1 in Fig.14.

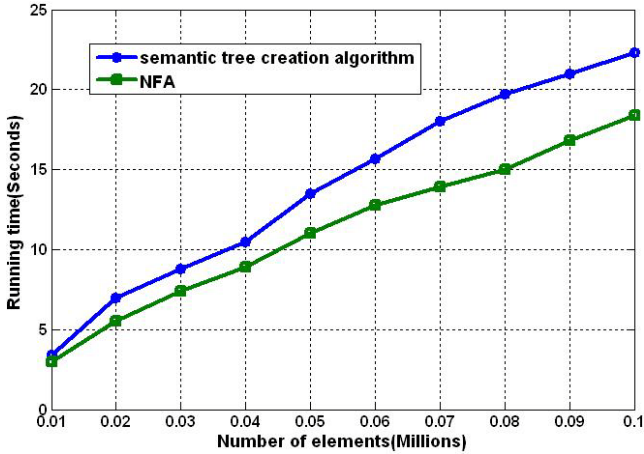

**Figure 14: The average running time between NFA and semantic tree creation algorithm over 100 thousand elements from Dataset1 based on the queries from Q11 to Q18 in Table 7.**

The idea of NFA is to filter the relevant elements with respect to keywords in order to reduce the running time of semantic tree creation algorithm [30]. It accesses all elements in a collection to get the semantic information similarity between the text of elements and keywords. Fig.14 shows that the semantic information of namespace in elements significantly reduce running time compared with semantic tree creation algorithm over a large collection of elements.

## 5. Related work

We have reviewed the literature of XML keyword search related to ours. To be the best of our knowledge, no existing work has formally studied on the namespaces [32] for elements retrieval. There has been a large body of work on content-oriented of XML documents and corresponding ranking schema.

Substantial researches have been done on the area of taking relevant matches between the content and the query as the criteria. e.g., DIL Query processing algorithm [4],Termjoin Algorithm [7] and Top-k algorithm [22]. Jovan Pehcevski et al. [10] content that purpose of XML retrieval task is to find elements that contain as much relevant information. However, some elements that are not keyword matches may be also relevant to the query but not return in those Algorithms. The classical method is to calculate the value of consine similarity between the content and keyword utilizing the formula of TF-IEF, the related work have been proposed in [6,17,18,19,23,26,27]. Unfortunately, most of them still cannot accurately calculate the similarity on semantic problem only by only this formula. Li Deng et al. [30] present the semantic tree creation algorithm. Other proposals are raised on semantic problem from the inner structure of XML document(e.g., Hongzhi Wang et al [13]; Norbert Govert et al [16]; Felix Weigel et al [20]; Sihem Amer-Yahia et al [22]; M.S.Ali et al [28]). However, they have to be faced with a large time execution. Benny Kimelfeld et al [15] have observed this shortcoming. They presented the method which filters the relevant documents before processing the Algorithm. Due to the notion of methods [15], we interestingly find that the namespace in elements not only solve the latent semantic problems between elements and keyword, but also filter the relevant elements based on the keyword to reduce time execution in the traditional algorithm. The most related work to this paper is [6,17], both of which have proposed the content of markup or frequency of markup as a factor contributed to semantic similarity between the content and query. However, It cannot effectively distinguish the elements with same markup representing different semantic information.

Another related area in elements retrieval is ranking schema based on keyword search. The classical scoring function is tf-ief(e.g., [12,22]) in information retrieval. However, many approaches simply calculate $tf_{t,e}$ with respect to all elements of the collection [9] or partly consider it by estimating $tf_{t,e}$ across elements of the same type [21]. $tf_{t,e}$ is also calculated based on the concentration of the text of the element and that of its descendants [14,21]. A different approach is to compute $tf_{t,e}$ for leaf-elements only, which are then used to score the leaf-elements themselves. All non-leaf elements are scored based on combination of the score of their descendants elements. The propagation of score starts from the leaf elements and can consider the distance between the element being considered and its descendent leaf-elements [24]. Similar notion is adopted by the DIL algorithm [4]. V.Mihajlovi et al. [25] ranks elements using a utility function that is based not only on the relevance score of an element, but also its size.

# 6.    Conclusion

This paper addresses the keyword search over elements in XML documents. Using the namespaces of elements, we have presented the Namespace Filter Algorithm (NFA) that retrieves the relevant components of XML documents with respect to keyword queries. In addition, we provide a new approach that can remove effectively the element overlaps occurring in query results. Using an evaluation formula, our approach is able to produce a ranking result-list without element overlaps. Compared with previous algorithms, NFA has demonstrated a better performance not only on time execution and but also on the precision and recall of query results. Our future work will study the relation on the previous factors on the background of graph structure in XML documents.

# 7.    Acknowledgements

We are grateful to thank the anonymous reviewers for their helpful comments.